\documentclass{aastex}
\usepackage{emulateapj5}
\usepackage{onecolfloat}
\usepackage{apjfonts}
\usepackage{subfigure}	

\begin{document}
\twocolumn[%

\title{High-Resolution Simulations of Cosmic Microwave Background 
non-Gaussian Maps in Spherical Coordinates} 

\author{M. Liguori\altaffilmark{1}, S. Matarrese\altaffilmark{1} and
L. Moscardini\altaffilmark{2}}

\begin{abstract}

We describe a new numerical algorithm to obtain high-resolution 
simulated maps of the Cosmic Microwave Background (CMB), for a 
broad class of non-Gaussian models. The kind of non-Gaussianity we  
account for is based on the simple idea that the primordial gravitational 
potential is obtained by a non-linear but local mapping 
from an underlying Gaussian random field, as resulting from a variety 
of inflationary models. Our technique, which is based on a direct 
realization of the potential in spherical coordinates and fully accounts for 
the radiation transfer function, allows to simulate non-Gaussian CMB maps 
down to the {\em Planck} resolution ($\ell_{\rm max} \sim 3,000$), 
with reasonable memory storage and computational time. 

\end{abstract}

\keywords{Cosmic microwave background --- Cosmology: theory --- statistics 
--- numerical simulations}
] 
\altaffiltext{1}{Dipartimento di Fisica `Galileo Galilei',
  Universit\'a di Padova, and INFN, Sezione di Padova, via Marzolo 8,
  I-35131 Padova, Italy; liguori@pd.infn.it, matarrese@pd.infn.it.}
\altaffiltext{2}{Dipartimento di Astronomia, 
  Universit\'a di Bologna, via Ranzani 1,
  I-40127 Bologna, Italy; moscardini@bo.astro.it.}

\section{Introduction}

It has been only recently realized that a certain degree of non-Gaussianity 
is a general prediction of all inflationary models for the generation of
primordial cosmological perturbations. Since the statistics of temperature
anisotropies of the Cosmic Microwave Background (CMB) directly probes that 
of the primordial fluctuations, any detection of non-Gaussianity in the CMB 
sky would provide a crucial test of the inflationary paradigm.  

The predicted degree of primordial non-Gaussianity is, however, generally
small; moreover, foreground contamination, instrumental noise 
and many secondary sources of anisotropy also generate 
non-Gaussian features in the CMB. Cosmic variance further complicates 
the problem, as some non-Gaussianity unavoidably arises from the uniqueness 
of the observed CMB sky (Scaramella \& Vittorio 1991; Srednicki 1993; 
Gangui {\em et al.} 1994; Gangui \& Martin 2000; Komatsu \& Spergel 
2001). All these effects make the search for non-Gaussianity a challenging 
observational and statistical task. 

Because of these facts, the quest for primordial non-Gaussian signals in the 
CMB requires high-resolution all-sky measurements, such as those 
recently achieved by the {\em Wilkinson Microwave Anisotropy Probe 
(WMAP)\/}\footnote{See http://map.gsfc.nasa.gov.} satellite 
(e.g. Spergel {\em et al.} 2003) and those that the forthcoming {\em Planck 
Surveyor\/}\footnote{See http://astro.estec.esa.nl/Planck.} will provide. 
At the same time one will need to develop optimal statistical 
estimators specifically designed to search for non-Gaussianity of a given 
type. This makes extremely important testing the power of different 
estimators on simulated all-sky CMB maps, with resolution comparable to that 
of the {\em WMAP} and {\em Planck} experiments. 
The aim of this paper is precisely that of providing an efficient numerical 
algorithm able to produce simulated CMB maps for a broad class of 
physically motivated non-Gaussian models. Simulations of non-Gaussian 
maps of the CMB sky have been recently performed with a number of different
techniques (Contaldi \& Magueijo 2001; Vio {\em et al.} 2002; 
Mart\'{i}nez-Gonz\'alez {\em et al.} 2002; Komatsu {\em et al.} 2003).  
The kind of models that we consider (see also Komatsu {\em et al.} 2003)
are based on the simple idea that the primordial (i.e. before being processed 
by the post-recombination radiation transfer function) gravitational 
potential $\Phi({\bf x})$ [actually Bardeen's gauge invariant variable 
$\Phi_H$ (Bardeen 1980)] can be written as a {\em local} but 
{\em non-linear} mapping from a `linear' random field 
$\Phi^{\rm L}({\bf x})$, which is Gaussian distributed with 
zero mean, namely 
\begin{equation}
\label{eqn:general}
\Phi({\bf x}) = {\cal F}[\Phi^{\rm L}({\bf x})] - 
\langle {\cal F}[\Phi^{\rm L}]\rangle \;,
\end{equation} 
where the last term on the RHS, has been added to ensure that $\Phi$ 
has zero mean, and only affects the mean temperature of the CMB.  
Models of this type were originally proposed by Coles and Barrow (1987), 
in connection with the statistics of CMB anisotropies in the large-angle 
(Sachs-Wolfe) limit. Moscardini {\em et al.} (1991) 
considered various models belonging to this class, as initial conditions for 
N-body simulations of large-scale structure formation in a Cold Dark 
Matter (CDM) cosmology. Falk, Rangarajan and Srednicki (1993) showed that 
perturbation produced during inflation, in single-field 
models, are characterized by a low non-Gaussianity level, 
which can be parametrized as (e.g. Wang \& Kamionkowski 2000) 
\begin{equation} 
\label{eqn:ourmodel} 
\Phi({\bf x}) = \Phi^{\rm L}({\bf x}) + f_{\rm NL} 
\left(\Phi^{\rm L}({\bf x})^2 - \langle 
(\Phi^{\rm L})^2 \rangle\right) \;, 
\end{equation}
where the dimensionless `non-linearity' parameter $f_{\rm NL}$
is much smaller than unity in their calculation, as a direct consequence of 
the tiny inflaton self-coupling. Similar conclusions were reached by Gangui 
{\em et al.} (1994), who used the techniques of 
stochastic inflation, to account 
for the back-reaction of field fluctuations on the background metric.
Related analyses have been made by Salopek, Bond \& Bardeen (1989), 
Salopek \& Bond (1990, 1991) and Wang \& Kamionkowski (2000), 
who accounted for possible features in the inflaton potential; 
Gupta {\em et al.} (2002), extended the stochastic calculation to
`warm-inflation' models, to evaluate the expected amplitude of 
non-Gaussianity.

The key idea in this field is to adopt the model described by 
Eq.(\ref{eqn:ourmodel}) as a prototype of the more general class of 
Eq.(\ref{eqn:general}), once the functional ${\cal F}$ is Taylor expanded 
beyond the linear term, as one can always do, given that only a mild level of 
non-Gaussianity is allowed by the data. 
Verde {\em et al.} (2000) showed that the optimal strategy to constrain 
this kind of models is to use (higher-order) statistics of the CMB, 
while Matarrese, Verde \& Jimenez (2000) and Verde {\em et al.} (2001) 
considered this class of models in connection with the abundance of 
high-redshift objects, which can be mapped to rare events of the underlying 
PDF (see also Komatsu {\em et al.} 2003). 

Quite recently, there has been a burst of interest for  
non-Gaussian perturbations of this type. Different CMB datasets have been 
analyzed, with a variety of statistical techniques (e.g. Komatsu 
{\em et al.} 2002; Ca\'yon {\em et al.} 2003; Santos {\em et al.} 2003), 
with the aim of constraining the `non-linearity' parameter $f_{\rm NL}$. 
The most stringent limit to date has been obtained by the {\em WMAP} team 
(Komatsu {\em et al.} 2003): $-58 <f_{\rm NL}< 134$ at $95~\%$ cl. 
Komatsu \& Spergel (2001) showed that the minimum value of 
$|f_{\rm NL}|$ which can be in principle detected using  
the angular bispectrum, is around 20 for {\em WMAP}, 
5 for {\em Planck} and 3 for an {\em ideal} experiment, owing to the intrinsic 
limitations caused by cosmic variance. Alternative strategies, 
based on the multivariate empirical distribution function of the spherical 
harmonics of a CMB map, have also been proposed 
(Hansen {\em et al.} 2002; Hansen, Marinucci \& Vittorio 2003). 

On the theoretical side, a massive analytical effort 
has been recently made in order to obtain a quantitative prediction for the 
non-linearity parameter in slow-roll inflation models (Acquaviva {\em et al.} 
2002; Maldacena 2002). The resulting expression appears to be more 
complicated than previously thought, as the parameter $f_{\rm NL}$  
is replaced by a suitable convolution kernel   
${\cal K}$, namely 
\begin{equation} 
f_{\rm NL} \Phi^{\rm L}({\bf x})^2 \to \int \!d^3 x_1\,d^3 x_2 
{\cal K}({\bf x}_1 - {\bf x},{\bf x}_2 - {\bf x})
\Phi^{\rm L}({\bf x}_1)\,\Phi^{\rm L}({\bf x}_2) \;.  
\end{equation} 
Although the `shape dependence' implied by the kernel (Acquaviva {\em et al.} 
2002; Maldacena 2002) might be an interesting way to look for specific 
inflationary non-Gaussian signatures, the typical values assumed by 
${\cal K}$ are of order $10^{-1} - 10^{-2}$, which makes the detection of 
non-Gaussianity produced in single-field slow-roll inflation a very 
challenging task! 
On the other hand, there are many physically motivated inflationary 
models, which can easily accommodate values of $f_{\rm NL}$ as large as, 
or even much larger than unity. 
This is the case, for instance, of a large class of multi-field inflation 
models, which leads to either non-Gaussian isocurvature perturbations 
(Linde \& Mukhanov 1997; Peebles 1997; Bucher \& Zhu 1997) or 
cross-correlated and non-Gaussian adiabatic and isocurvature 
modes, as first realized by Bartolo, Matarrese \& Riotto 
(2002) (see also Bernardeau \& Uzan 2002). Other interesting possibilities 
include the so-called {\em curvaton} model, where the late time decay of 
a scalar field, belonging to the non-inflatonic sector of the theory,  
induces curvature perturbations (e.g. Mollerach 1990; Lyth \& Wands 2002)
which are characterized by values of $f_{\rm NL} \gg 1$ (Lyth, Ungarelli \& 
Wands 2003). The possible role of local fluctuations in the inflaton coupling 
to ordinary matter in driving non-Gaussianity has been recently analyzed 
(Dvali, Gruzinov \& Zaldarriaga 2003; Kofman 2003; Zaldarriaga 2003). 
Non-Gaussianity caused by the effect of higher-dimension operators in 
the Lagrangian of single-field inflation models has also been considered 
(Creminelli 2003). 
   
The model considered here will be that of Eq.(\ref{eqn:ourmodel}), 
which, besides the advantage of its simplicity, is very 
useful to constrain the non-Gaussianity level in terms of a 
single parameter $f_{\rm NL}$. 

The plan of the paper is as follows. In Section 2 we discuss the 
various steps which lead to the construction of our simulation algorithm. 
Section 3 contains our main results and some concluding remarks. 

\section{Outline of the method}

\begin{figure}[!t] 
\begin{center} 
\includegraphics[height=0.45\textheight,width=0.5\textwidth]{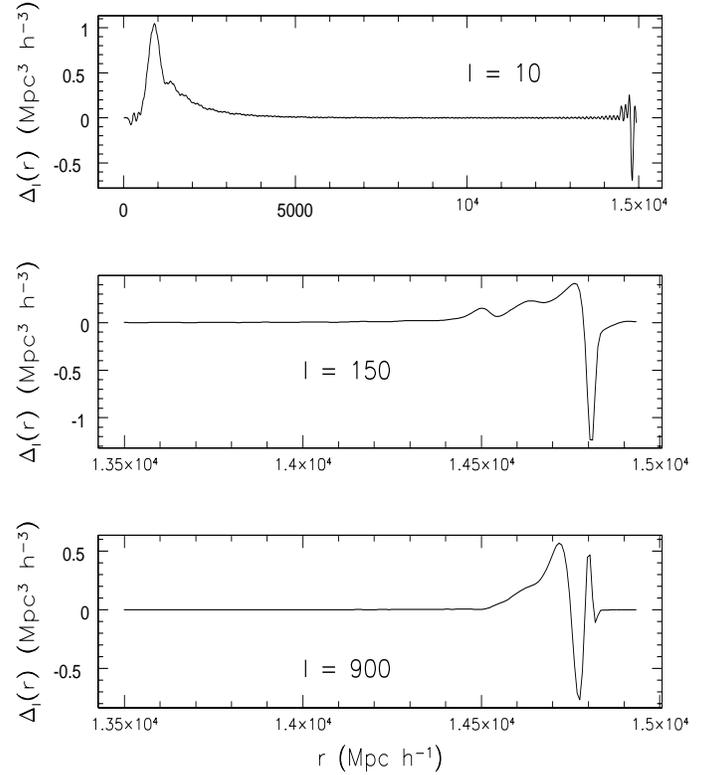} 
\caption{The coefficients $\Delta_\ell(r)$ evaluated for different values 
of $\ell$, assuming a $\Lambda$CDM cosmology ($h=0.65$, $\Omega_b = 0.05$, 
$\Omega_c=0.25$, $\Omega_\Lambda=0.7$). We put $r=c\eta$, where $\eta$ is 
the conformal time. In the chosen model, $c\eta_0 \sim 14.9 Gpc$ 
and $c\eta_* \sim 289$ Mpc, where $\eta_0$ and $\eta_*$ 
are the present day and 
the recombination conformal time respectively.}\label{fig:deltar} 
\end{center}
\end{figure}

The primordial potential fluctuations can be related to temperature 
anisotropies of the CMB by convolution with the 
{\em radiation transfer function} $\Delta_\ell(k)$,    
\begin{equation}
\label{eqn:transfer} 
a_{\ell m} = 4 \pi (-i)^\ell 
\int\!\frac{d^3k}{(2\pi)^3} \,\,\Phi(\mathbf{k})\,\Delta_\ell(k)\,
Y^*_{\ell m}(\hat{k}) \; , 
\end{equation} 
where the temperature fluctuation in the direction $\hat n$ has been 
expanded in spherical harmonics $Y_{\ell m}({\hat n})$ with multipole 
coefficients $a_{\ell m}$, namely 
$\frac{\Delta T}{T} ({\hat n}) = \sum_{\ell m} a_{\ell m} 
Y_{\ell m}({\hat n})$.  
The problem analyzed here is how to simulate non-Gaussian 
temperature maps starting from the non-linear primordial gravitational 
potential described by Eq. (\ref{eqn:ourmodel}). 
To this aim it is necessary to obtain the non-Gaussian part of the potential 
as a function of the Gaussian one, which can be generated by standard
techniques, and finally account for the radiative transfer, in 
order to determine the Gaussian and non-Gaussian contributions to 
the temperature harmonic coefficients $a_{\ell m}$. 
At face value the simplest approach seems to make realizations of the 
primordial potential in Fourier space and then replace it directly in 
equation (\ref{eqn:transfer}). However, this approach presents some technical
drawbacks, as we are going to explain. In Fourier space, the non-Gaussian
part of the potential is a convolution product of Gaussian fields (e.g.
Komatsu \& Spergel 2001): 
\begin{equation}
\label{eqn:convolution}
\Phi^{\rm NL}(\mathbf{k}) = f_{\rm NL}\int \!\frac{d^3p}{(2\pi)^3}
\,\,\Phi^{\rm L}(\mathbf{k+p})\, \Phi^{\rm L}(\mathbf{p}) \; , 
\end{equation} 
so, in principle we may obtain $\Phi^{\rm NL}(\mathbf{k})$ by a numerical 
evaluation of the convolution, by starting from a Gaussian potential 
field with given power-spectrum, generated on a Fourier-space grid. 
The computational problems in this procedure arise from the size of the 
simulation box.  
We have in fact to consider a box-size of the order of the present day cosmic 
horizon, $L_{\rm box} = 2c\eta_0 \sim 20 \; h^{-1}$ Gpc, 
(where $\eta_0$ is the value of the conformal time today) and we need a 
resolution of about $20 h^{-1}$ Mpc to accurately resolve the last scattering 
surface (e.g Komatsu {\em et al.} 2003; see also Figure 1): 
from this we find that, in order to generate the Gaussian part 
of the potential we need at least, $1024^3$ points on the grid.

\footnote{Note 
that at any stage one can replace the square operation by whatever local 
functional of $\Phi_{\rm L}$, with no extra computational complications.}.

So, the recipe for the generation of the non-Gaussian maps could 
be the following: {\em i)} generate a Gaussian potential in Fourier space, 
using $1024^3$ grid-points; 
{\em ii)} go to real space by inverse Fast Fourier Transform (FFT), and 
square it to obtain the non-Gaussian contribution\footnote{Note that one can 
replace the square operation by whatever local functional of 
$\Phi_{\rm L}$, with no extra computational complications.}; {\em iii)} 
transform back to Fourier space and convolve the harmonic transforms of the 
Gaussian and non-Gaussian parts [see Eq. (\ref{eqn:transfer})] with the 
radiation transfer function $\Delta_{\ell}(k)$, to get the CMB multipoles. 
If we call $\Phi_{\ell m}(k)$ the multipoles of the harmonic expansion of 
the primordial gravitational potential, the CMB multipoles are in fact 
given by  
\begin{equation}
\label{eqn:transfer2} 
a_{\ell m} =
\frac{(-i)^{\ell}}{2\pi^2} \int \! dk \,\, k^2 \, \Phi_{\ell m}(k) \,
\Delta_\ell(k) \; , 
\end{equation} 
where $\Phi_{\ell m}(k)  \equiv \Phi_{\ell m}^{\rm L}(k) + f_{\rm NL} 
\Phi_{\ell m}^{\rm NL}(k)$ and, for each term,   
\begin{equation} 
\label{eqn:multipoles_definitions} 
\Phi_{\ell m}(k) \equiv \int \! d\Omega_{\hat{k}} \, 
\Phi(\mathbf{k}) \, Y_{\ell m}(\hat{k}) \; . 
\end{equation} 

The method we have just outlined can be easily improved by accounting for 
radiative transfer in real rather than Fourier space: to see this 
let us start from the harmonic transform of the primordial potential in real 
space and call it $\Phi_{\ell m}(r)$ [in analogy with the previous definitions
we will also introduce the quantities $\Phi_{\ell m}^{\rm L}(r)$ and 
$\Phi_{\ell m}^{\rm NL}(r)$]. We can find the relation between any pair of
$\Phi_{\ell m}(k)$ and $\Phi_{\ell m}(r)$ (see the Appendix), namely  
\begin{equation}
\label{eqn:ak2ar} 
\Phi_{\ell m}(r) =
\frac{(-i)^{\ell}}{2 \pi^2} \int \! dk \, k^2 \, \Phi_{\ell m}(k) \,
j_\ell(kr) \; , 
\end{equation} 
and its inverse 
\begin{equation}
\label{eqn:ar2ak} 
\Phi_{\ell m}(k) = 4 \pi (i)^\ell \int
\! dr \, r^2 \, \Phi_{\ell m}(r) \, j_\ell(kr) \; , 
\end{equation} 
where $j_\ell$ is the spherical Bessel function of order $\ell$.

If we now replace Eq. (\ref{eqn:ar2ak}) in Eq. (\ref{eqn:transfer2}), 
exchange the order of integrations and define

\begin{equation}
\label{eqn:rtf_definition} 
\Delta_{\ell}(r) \equiv
\frac{2}{\pi} \int \! dk \, k^2 \, \Delta_\ell(k) j_\ell(kr) \; ,
\end{equation} 
we can finally write:  
\begin{equation}
\label{eqn:rtf}
a_{\ell m} = \int \! dr \, r^2 \Phi_{\ell m}(r) \Delta_{\ell}(r) \;. 
\end{equation}
 
The numerical advantages deriving from Eq. (\ref{eqn:rtf}) are clear: 
if we generate CMB multipoles by this equation we can pre-compute and store 
the quantities $\Delta_{\ell}(r)$ for all the simulations characterized by 
the same cosmological model and we have to FFT only once, instead of going 
to real space and then back to Fourier space. The numerical calculation of the
$\Delta_{\ell}(r)$ has been obtained by a modification of the CMBfast
code (Seljak and Zaldarriaga 1996); the behaviour for some of these 
coefficients, evaluated for a $\Lambda$CDM cosmology, is shown in Figure 1. 
  
To summarize: a numerically efficient technique for the 
generation of non-Gaussian maps is to generate a Gaussian  
field on a grid in Fourier space, then FFT and square it to get the 
non-Gaussian part of the gravitational potential in real space and finally 
harmonic transform and convolve the two contributions with $\Delta_\ell(r)$ 
to obtain Gaussian $a_{\ell m}^{\rm L}(r)$ and non-Gaussian ones, 
$a_{\ell m}^{\rm NL}(r)$.
For any value of $f_{\rm NL}$ the CMB multipoles will then be given by  
$a_{\ell m} = a_{\ell m}^{\rm L} + f_{\rm NL}\,a_{\ell m}^{\rm NL}$. 

\begin{figure}[!t] 
\begin{center} 
\includegraphics[height=0.45\textheight,width=0.5\textwidth]{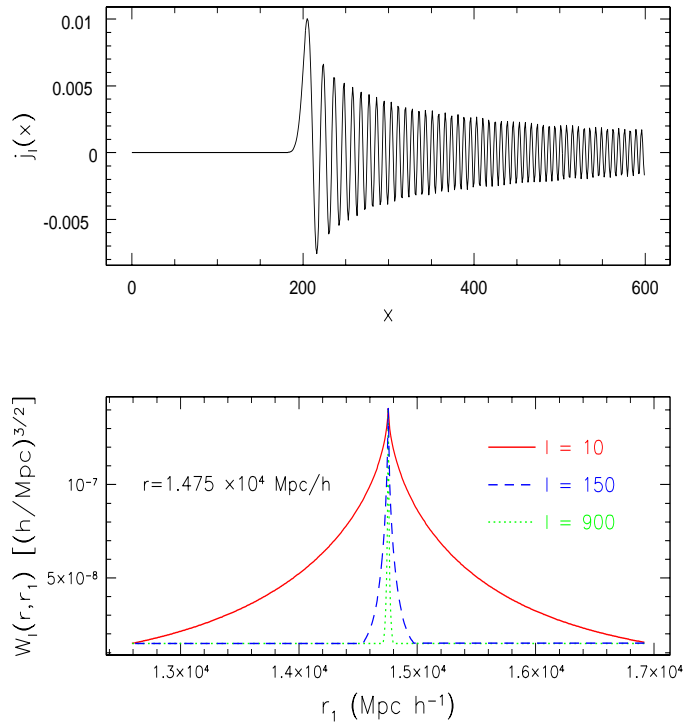} 
\caption{The top panel shows the spherical Bessel function of order $150$. 
The bottom panel shows the filters $W_\ell(r,r_1)$ as functions of $r_1$, at
fixed $r$, and three different values of $\ell$. We choose the spectral index 
$n=1$. Note that the filters become more and more peaked around $r_1 = r$ as 
$\ell$ increases. The Bessel functions 
oscillate very fast and have to be sampled in many more points than $W_\ell$
to have an accurate numerical integration. For this reason the use of $W_\ell$
instead of $j_\ell$ drastically reduces the needed CPU time, as explained
in the text.}\label{fig:Wl} 
\end{center}
\end{figure}

Komatsu {\em et al.} (2003) (see also Komatsu, Spergel \& Wandelt 2003)
used this method to generate $300$ non-Gaussian maps at the {\em WMAP} 
resolution; their algorithm required about $3$ hours on $1$ processor of 
SGI Origin $300$ and $7.5$ GB of physical memory to generate a single map.\\

We developed an alternative algorithm which starts from the 
generation of the linear potential multipoles 
$\Phi^{\rm L}_{\ell m}(r)$, while we use the same technique as 
Komatsu {\em et al.} (2003) to account for the radiative transfer. 
We believe that the main advantages of the method we are going to describe 
w.r.t. the one just outlined can be the following: 
{\em i)} much less memory requirements, 
{\em ii)} a decrease of CPU time, 
{\em iii}) the possibility to more accurately resolve the last-scattering 
surface.  

As we are going to see, we will obtain the multipole coefficients 
$\Phi^{\rm L}_{\ell m}(r)$ directly, i.e. without passing through the 
generation of the potential on a cubic grid. 
So we won't need to convert from Cartesian to spherical coordinates, in 
order to harmonic transform $\Phi^{\rm L}(\mathbf{r})$ and, even more 
important, we won't need to store big arrays containing the values of 
$\Phi^{\rm L} (\mathbf{r})$, calculated on the grid points. 
As the $\Phi^{\rm L}_{\ell m}(r)$ are Gaussian
variables, to generate them we only need to know their correlation
function. We have (see the Appendix): 
\begin{equation}
\label{eqn:almrcorr} \left \langle \Phi^{\rm L}_{\ell_1
m_1}(r_1)\Phi^{{\rm L}\star}_{\ell_2 m_2}(r_2) \right \rangle = \frac{2}{\pi}
\delta_{\ell_1}^{\ell_2} \delta_{m_1}^{m_2} \int \!dk\,k^2 P_\Phi(k)
j_{\ell_1}(kr_1) j_{\ell_2}(kr_2) \; , 
\end{equation} 
where $P_\Phi(k)$ is the primordial (i.e. unprocesssed by the radiation 
transfer function) power-spectrum of the gravitational potential, and 
$\delta^i_j$ is Kronecker's delta. 
Since the multipoles $\Phi^{\rm L}_{\ell m}$ are correlated in real space, 
they cannot be generated directly. One possibility is to generate 
$\Phi^{\rm L}_{\ell m}(k)$ first, as they are independent Gaussian variables 
(see the Appendix, for an explicit calculation),
\begin{equation}
\label{eqn:almkcorr} 
\left \langle 
\Phi^{\rm L}_{\ell_1 m_1}(k_1) \Phi^{{\rm L}\star}_{\ell_2 m_2}(k_2) 
\right \rangle = 8\pi^3
\frac{P_\Phi(k_1)}{k_1^2} \delta^D(k_1-k_2)\delta_{\ell_1}^{\ell_2}
\delta_{m_1}^{m_2} \; , 
\end{equation} 
($\delta^D$ being the Dirac delta function) and then transform 
$\Phi_{\ell m}(k)$ into $\Phi_{\ell m}(r)$, by Eq. (\ref{eqn:ak2ar}).\\ 

As in this last equation, $\ell$ and $m$ are fixed and only $k$ varies for 
each integration, it is clear that this method does not require large memory: 
in fact, we never need to store all the values of 
$\Phi_{\ell m}(k)$ and $\Phi_{\ell m}(r)$ at the same time in physical 
memory.\\ 
Unfortunately, the spherical Bessel transform which allows the passage from 
$\Phi_{\ell m}(k)$ to $\Phi_{\ell m}(r)$ requires a lot of CPU time, 
as the $j_\ell(kr)$ are rapidly oscillating functions which have to be sampled
in many points to ensure accurate numerical integration. 
The time needed to simulate one sky-map with this method is  
around $20$ hours on a Compaq server DS20E ($500$ Mhz), for $l_{max}=500$, 
and scales roughly
\footnote{This is because the total number of multipoles, and 
consequently the number of transforms to be calculated, is $\sim l^2$} 
as $l^2$. However, it is clear from the previous argument that the code could 
be greatly speeded up by avoiding the numerical integration of Eq.
(\ref{eqn:ak2ar}). This result can be achieved as follows: start from 
{\em independent} complex Gaussian variables $n_{\ell m}(r)$, 
characterized by the correlation function 
\begin{equation}
\label{eqn:whitenoise} 
\left \langle n_{\ell_1 m_1}(r_1)
n^*_{\ell_2 m_2}(r_2) \right \rangle = 
\frac{\delta^D(r_1-r_2)}{r^2}\delta_{\ell_1}^{\ell_2} \delta_{m_1}^{m_2}\; ;  
\end{equation} 
as shown in the Appendix, it is now possible to recover our Gaussian variables 
$\Phi^{\rm L}_{\ell m}(r)$, with the right correlation properties, by 
convolution, 
\begin{equation}
\label{eqn:real_convolution} 
\Phi^{\rm L}_{\ell m}(r) = \int \! dr_1 \, r_1^2 \, n_{\ell m}(r_1) 
W_\ell(r,r_1) \; ,
\end{equation} 
where the filter functions $W_\ell(r,r_1)$ are defined as
\begin{equation}
\label{eqn:filter} 
W_\ell(r,r_1) =
\frac{2}{\pi} \int \! dk \, k^2 \, \sqrt{P_\Phi(k)} \, j_\ell(kr)
j_\ell(kr_1) \; .  
\end{equation} 

In this work we will only consider models with a scale-free primordial 
density power-spectrum of the form $P(k) = Ak^n$, and we take $n=1$, in
agreement with observational data (e.g. Spergel {\em et al.} 2003). 
In this case, when $r$ is fixed, the functions $W_\ell(r,r_1)$ are smooth and 
differ from zero only in a narrow region around $r=r_1$  
(see Figure 2). So, in order to accurately sample $W_\ell$ for their 
integration in Eq. (\ref{eqn:real_convolution}) we only need to compute them 
in a few points, which considerably reduces the CPU time needed for the 
generation of $\Phi^{\rm L}_{\ell m}(r)$. 
The CPU time needed for the generation of a single sky-map
at the {\em WMAP} resolution now goes down to less than one hour. 
In other words, by the latter method 
we work directly in real space, starting from white-noise coefficients and 
convolving them with suitable filters to produce the right correlation
properties of the $\Phi^{\rm L}_{\ell m}(r)$. 
This technique for the generation of correlated random variables 
is well known in the literature (e.g. Salmon 1996; see also Peebles 1983). 
It is particularly efficient in our case, because, as it appears from Figure 2
the $W_\ell(r,r_1)$ get more and more peaked around $r=r_1$ as $\ell$ 
increases.

\section{Results and Conclusions}

Let us summarize the various steps which define our numerical algorithm.

\begin{itemize}
\item Simulate the white-noise Gaussian fields  
$n_{\ell m}(r)$ directly in real space.\\
\item Convolve them with the (pre-computed) filter
functions $W_\ell(r,r_1)$, to obtain the
linear potential multipoles $\Phi^{\rm L}_{\ell m}(r)$.\\
\item Compute the linear potential 
$\Phi^{\rm L}({\bf r}) = \sum_{\ell m} \Phi^{\rm L}_{\ell m}(r) Y_{\ell m}
({\hat r})$ and square it to obtain (modulo $f_{\rm NL}$) the non-Gaussian 
contribution to the gravitational potential, directly in spherical 
coordinates.\\
\item Harmonic transform\footnote{Harmonic transforms 
have been performed using the HEALPix package 
[http://www.eso.org/science/healpix/ see also (G\'orski, Hivon \& Wandelt 
1999)].} $\Phi^{\rm NL}({\bf r})$ to get $\Phi^{\rm NL}_{\ell m}(r)$.\\
\item Convolve $\Phi_{\ell m}(r) = 
\Phi^{\rm L}_{\ell m}(r) + f_{\rm NL}
\Phi^{\rm NL}_{\ell m}(r)$ with the (pre-computed) real-space radiation 
transfer functions $\Delta_\ell(r)$. 
\end{itemize} 

\begin{figure}[!t] 
\begin{center} 
\includegraphics[height=0.45\textheight,width=0.5\textwidth]{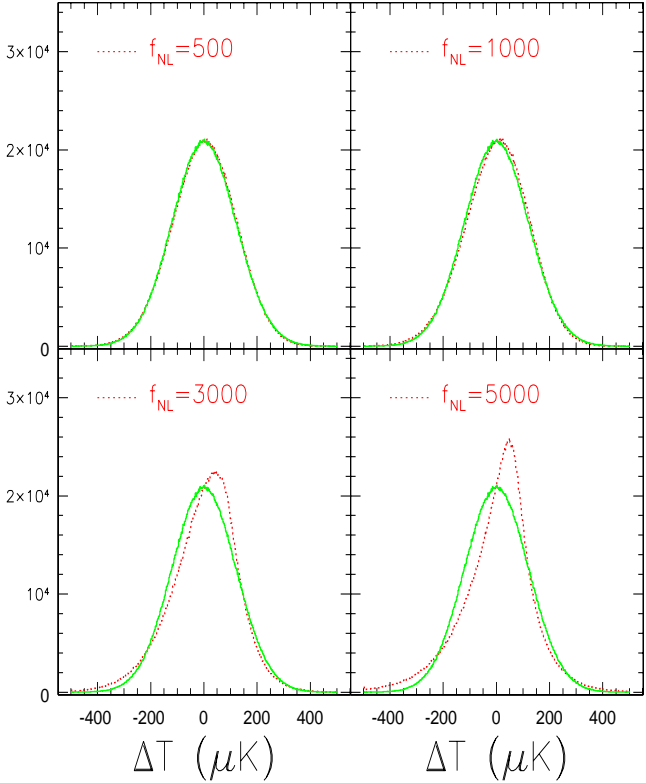} 
\caption{One-point PDF of temperature fluctuations obtained from a single
simulated map with $l_{max} = 3000$. Different panels show a comparison 
between the Gaussian realization ($f_{\rm NL}=0$) and the corresponding
non-Gaussian one, for different values of $f_{\rm NL}$. Beam smearing has not
been included here.}\label{fig:pdf} 
\end{center}
\end{figure}

The form of $\Delta_\ell(r)$
suggests to generate the potential coefficients $\Phi^L_{\ell m}(r)$ for 
non-uniformly spaced values of the radial coordinate. In the final
convolution, the real-space radiation transfer functions are in fact
different from zero only near last scattering ($0.1 \eta_* \lesssim \eta 
\lesssim 2 \eta_*$, see Figure \ref{fig:deltar}; 
$\eta_*$ is the conformal time,
correspondingly $r_* = c \eta_*$), with a further contribution, coming from 
low redshifts and multipoles 
($\ell \lesssim 50$, $r < 5000 \, h^{-1}$ Mpc, see 
the lower panel of Figure \ref{fig:deltar}), due to the late-ISW effect. 
Therefore, we generated the multipoles 
$\Phi^L_{\ell m}(r)$ only for $\ell$ and $r$ belonging to the previously 
defined intervals\footnote{Actually, at low redshifts we generate 
$\Phi^L_{\ell m}(r)$ for $\ell \leq 100$, even if we are
ultimately interested only in the multipoles with $\ell \leq 50$; 
this is necessary as high-order Gaussian multipoles contribute to 
lower-order non-Gaussian ones in the harmonic transforms.}. 
In this way we can increase the radial resolution 
in the selected regions and, at the same time, reduce the total number of
points in our spherical grid (i.e. reduce the total number of evaluated
$\Phi_{\ell m}(r)$), with a consistent reduction of the total CPU time needed. 
Moreover, since at low redshifts we need to consider only low multipoles, we 
may in this case reduce the pixel resolution when using HEALPix to perform
the harmonic transforms, with further reduction of CPU time 
(this is because the value of the parameter 
$nside$ in HEALPix, which is directly related to the pixel resolution, 
has to be chosen in such a way as to verify the relation  
$l_{\rm max} < 3 \times nside$, and the calculation time for the harmonic 
transforms scales as $(nside)^3$). 

In other words, our multipole approach enables to adapt the simulation grid
to the optimal sampling of the physically relevant radial intervals
(i.e. points where the radiative transfer is non-zero) and to reduce the 
CPU time needed for the generation of $\Phi^{L,NL}_{\ell m}(r)$. 
Moreover, we found it useful to split the high redshift contributions 
(SW effect, acoustic oscillations) 
from the low redshift contributions (late-ISW effect): in this way the 
simulation of the latter effects can be greatly speeded up, as they 
affect only the lower multipoles of the final map. 

We have applied our algorithm to obtain some simulated temperature maps with 
different values of $\ell_{max}$, starting from a $\Lambda$CDM model with
primordial spectral index $n=1$. The radial coordinate
has been discretized with a step of $7.5 \, h^{-1}$ Mpc in the regions
where $\Delta_\ell(r)\neq 0$. In this way we generated 
$100$ potential multipoles in the radial interval $(r_0 - 2r_*) 
\leq r \leq(r_0 - 0.1r_*)$, for each $\ell$,$m$, $2 \leq \ell 
\leq l_{\rm max}$, $0 \leq m \leq \ell$, and we generated 
about $600$ potential multipoles with
 $r \leq 5000 \, h^{-1}$ Mpc, for each $2 \leq \ell \leq 100$, 
 $0 \leq m \leq \ell$.
 
To run our simulations we need $200$ Mb of physical memory when
$\ell_{\rm max} = 3000$; the final CPU time  is reported in Table 
(\ref{tab:times}), where we also show the time needed for the
harmonic analysis, at large $r$ only, as this is the most time-consuming 
part of our algorithm. 

\begin{figure*}[p]
\centering

\subfigure{\includegraphics[height=0.2\textheight,width=0.4\textwidth]
{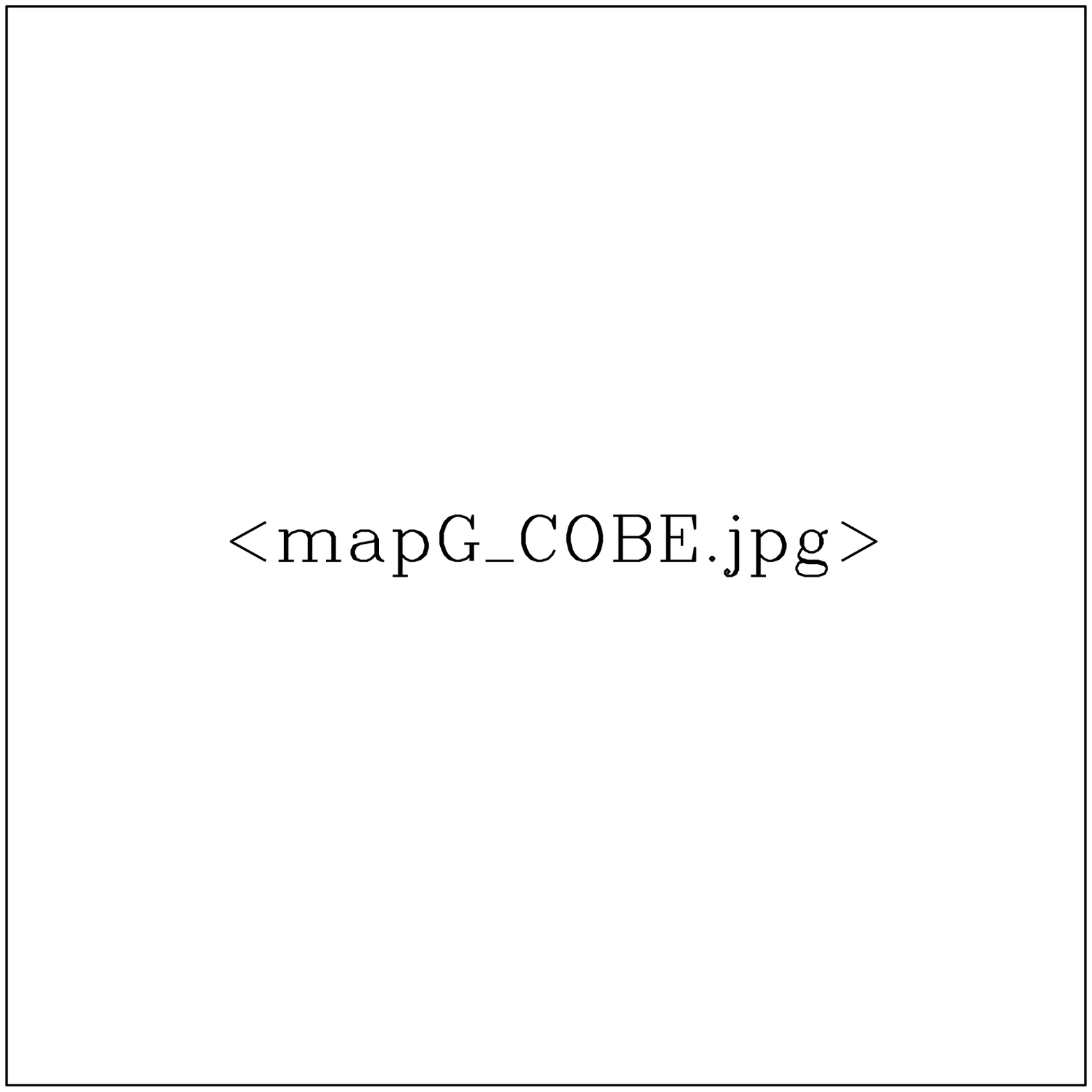}} \qquad
\subfigure{\includegraphics[height=0.2\textheight,width=0.4\textwidth]
{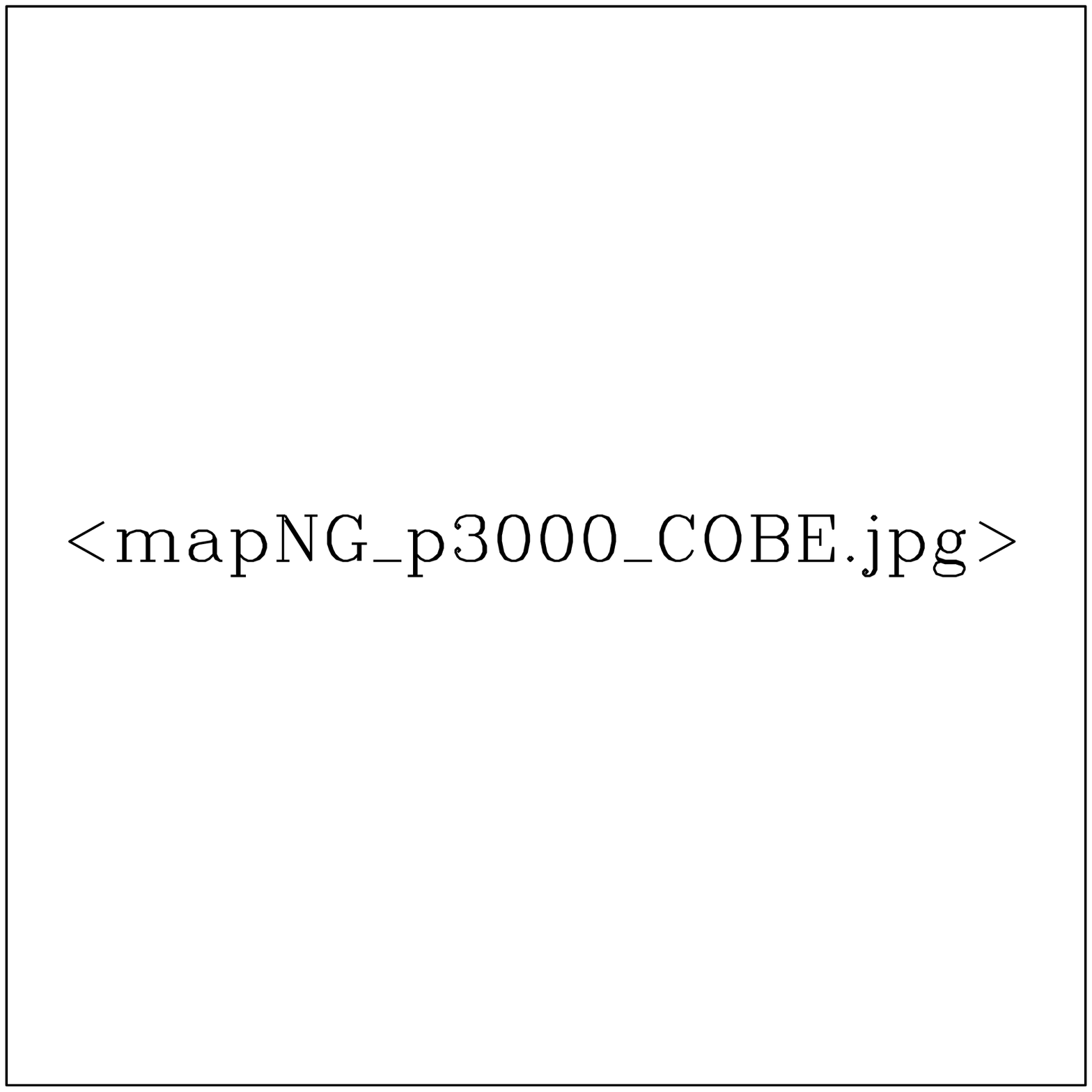}} \\

\subfigure{\includegraphics[height=0.2\textheight,width=0.4\textwidth]
{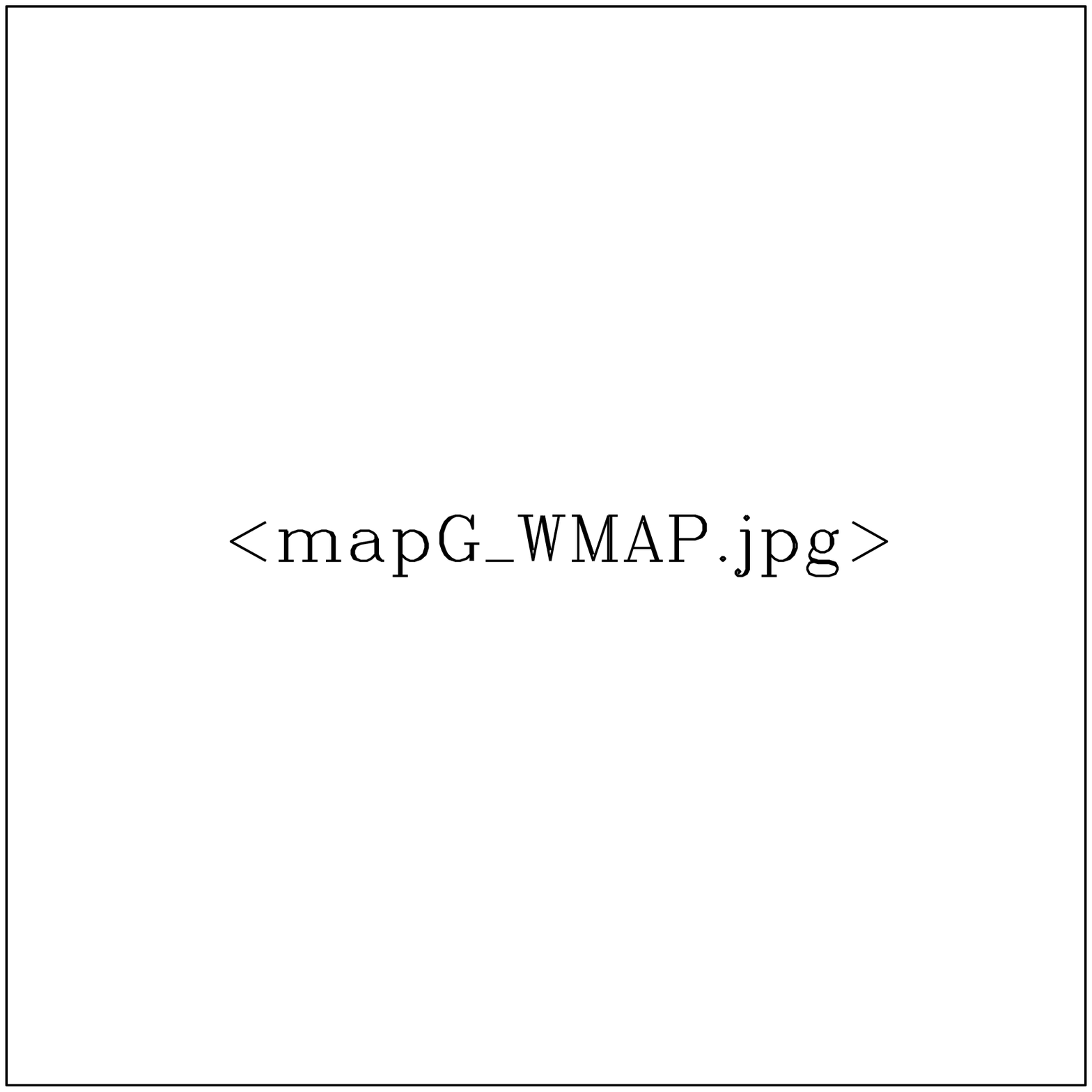}} \qquad
\subfigure{\includegraphics[height=0.2\textheight,width=0.4\textwidth]
{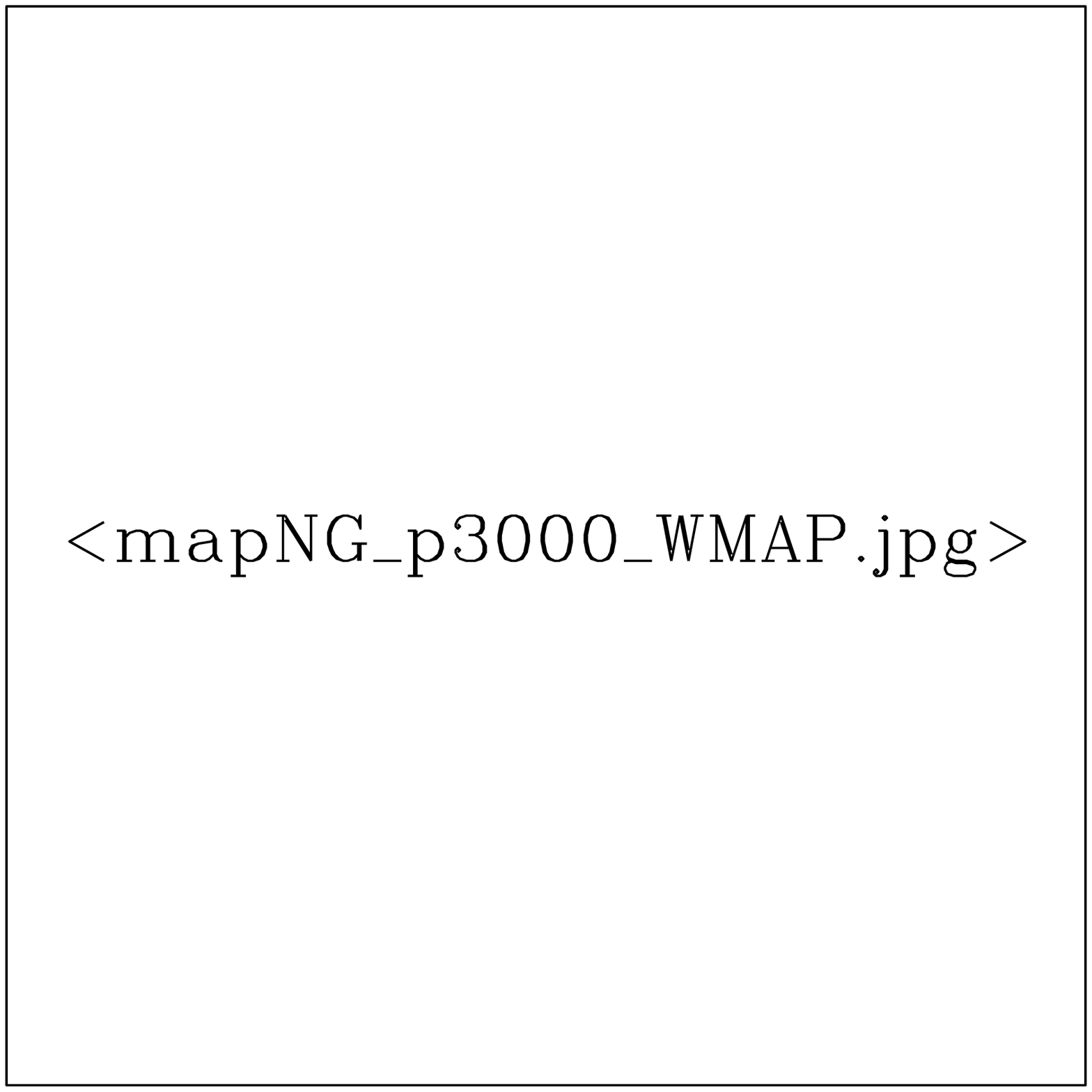}} \\

\subfigure{\includegraphics[height=0.2\textheight,width=0.4\textwidth]
{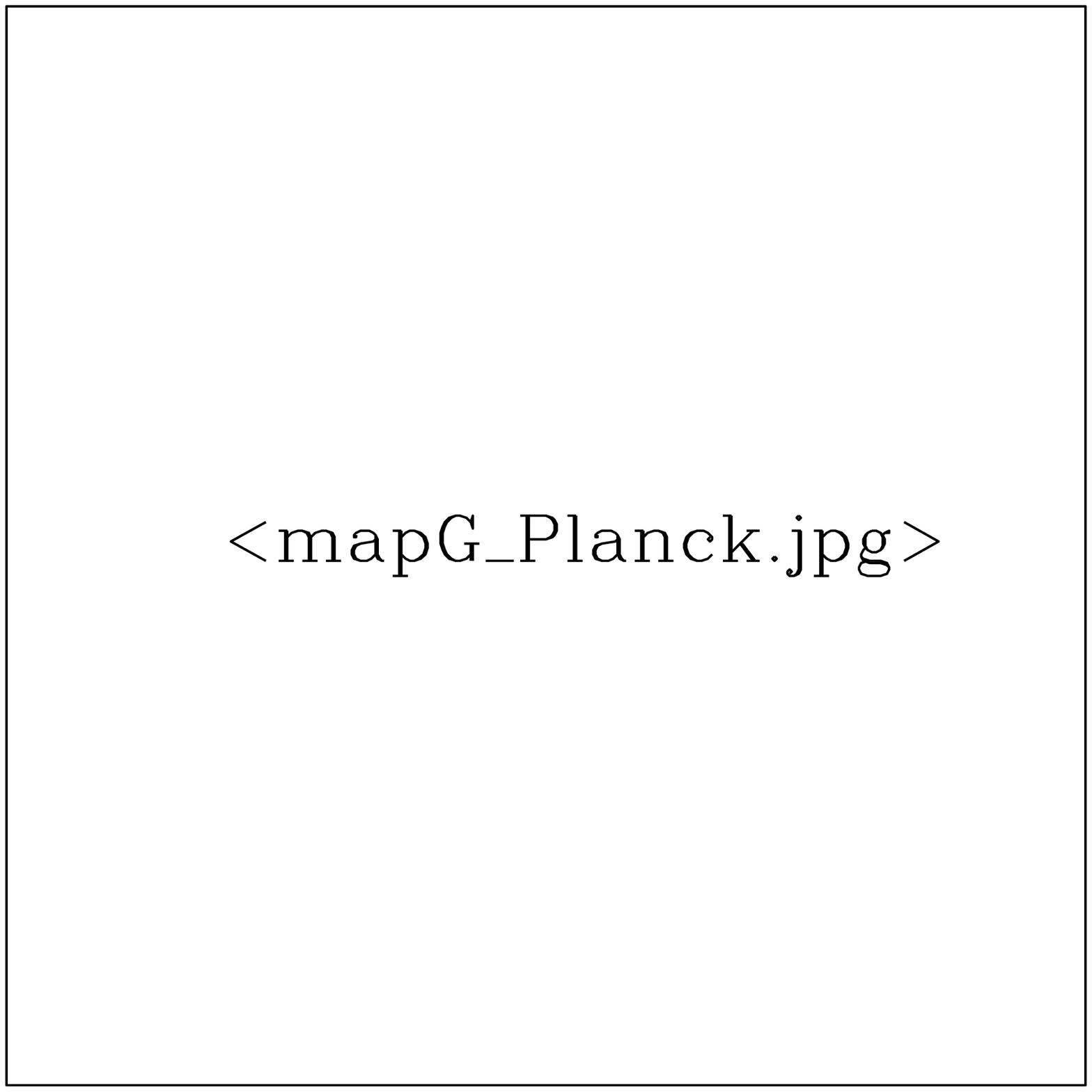}} \qquad
\subfigure{\includegraphics[height=0.2\textheight,width=0.4\textwidth]
{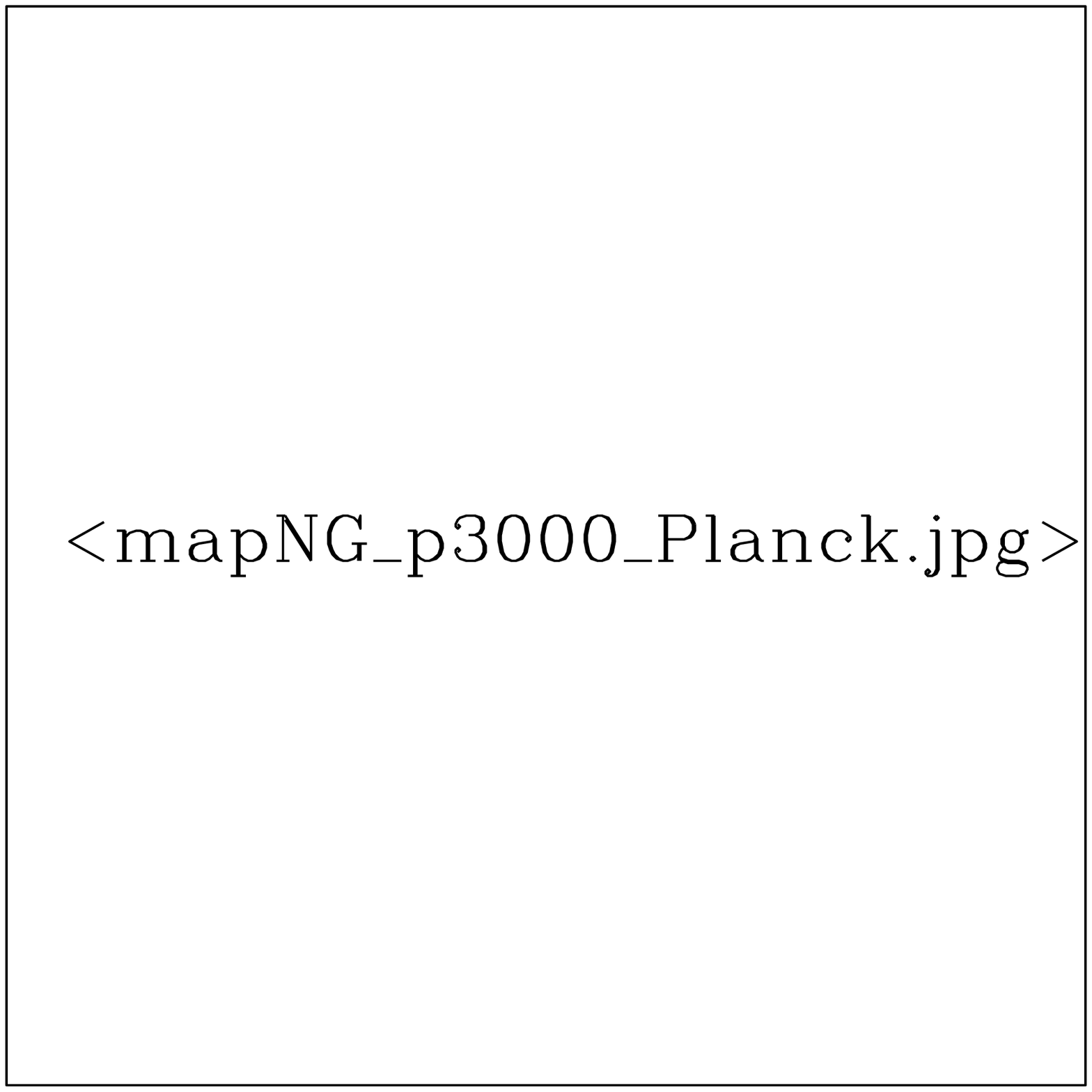}}

\caption{The left-hand side of this panel shows the same Gaussian 
realization, smoothed by three different beams. From top to bottom, 
the FWHM of the beams is $7^{\circ}$ ({\em{COBE}}), 13 arcmin 
({\em{WMAP}}) and 5 arcmin ({\em{Planck}}). The right-hand side shows three 
corresponding non-Gaussian realizations, obtained from the Gaussian one by 
adding a non-linear coupling parameter $f_{\rm NL} = 3000$ (such a high value 
of $f_{\rm NL}$ is chosen to make the non-Gaussian effects visible by eye). 
The model is a $\Lambda$CDM with primordial spectral index $n=1$. 
The primordial fluctuations have been {\em{COBE}}-normalized 
by CMBfast.}\label{fig:maps}    

\end{figure*}

\begin{figure*}[p]
\centering

\subfigure{\includegraphics[height=0.2\textheight,width=0.4\textwidth]
{G1.ps}} \qquad
\subfigure{\includegraphics[height=0.2\textheight,width=0.4\textwidth]
{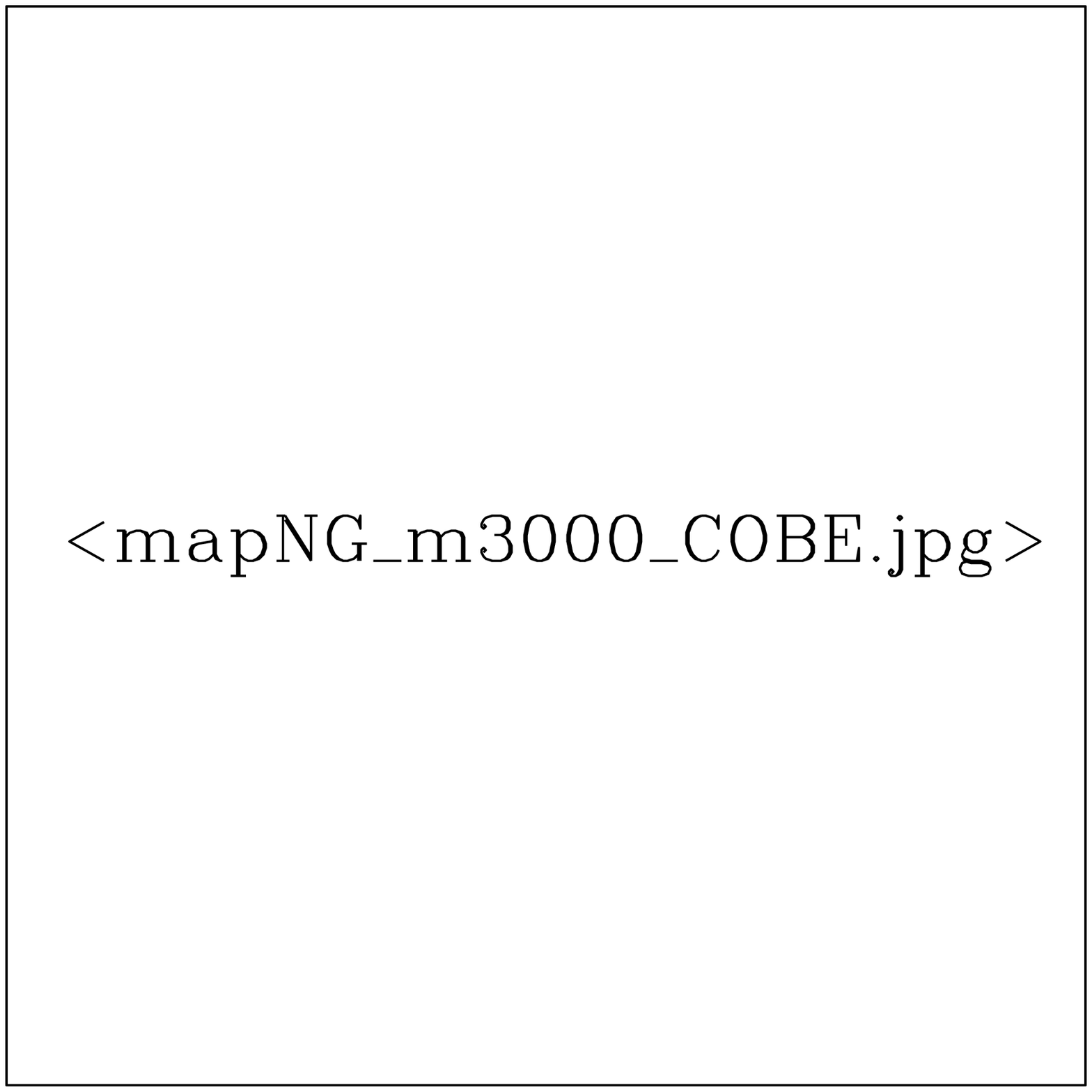}} \\

\subfigure{\includegraphics[height=0.2\textheight,width=0.4\textwidth]
{G2.ps}} \qquad
\subfigure{\includegraphics[height=0.2\textheight,width=0.4\textwidth]
{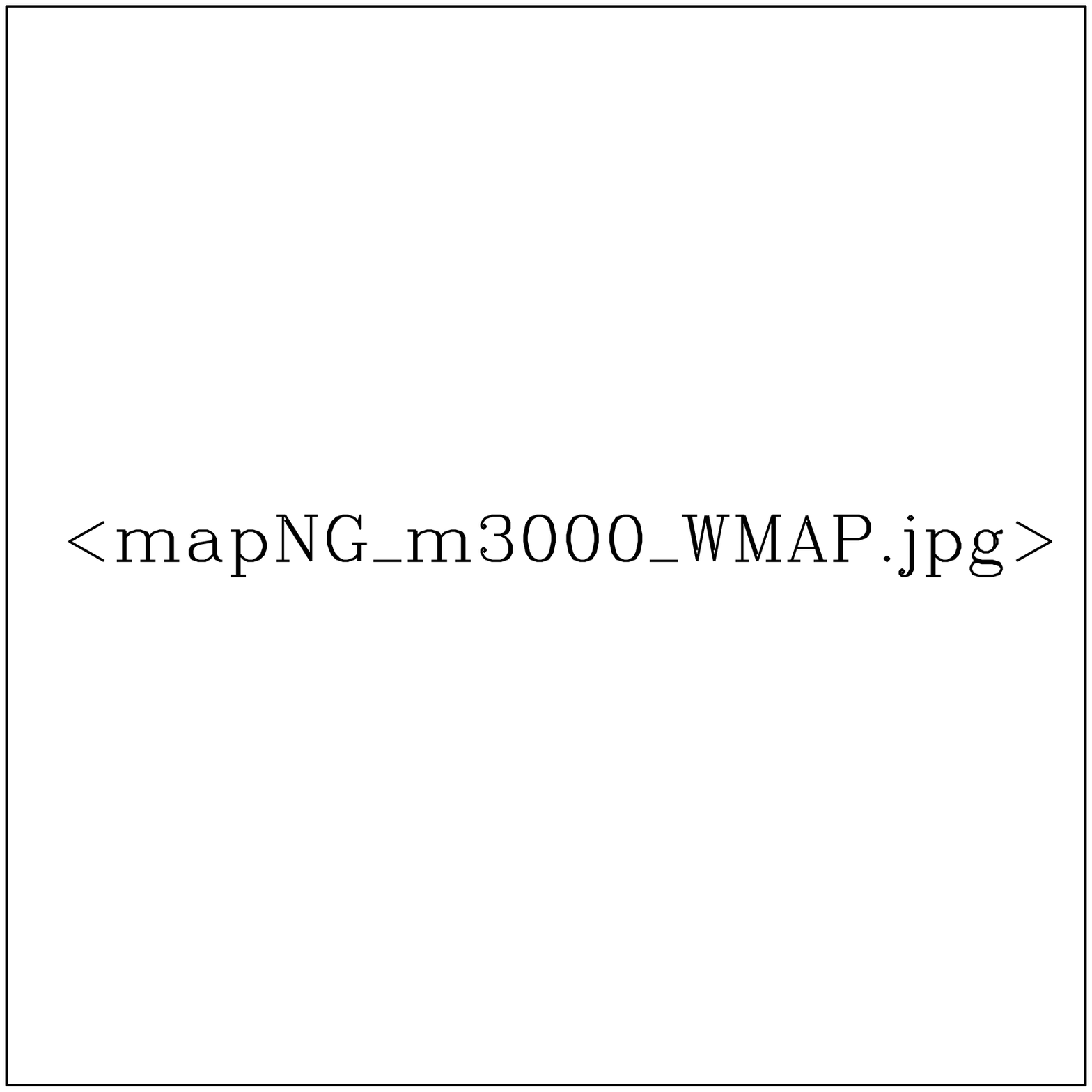}} \\

\subfigure{\includegraphics[height=0.2\textheight,width=0.4\textwidth]
{G3.ps}} \qquad
\subfigure{\includegraphics[height=0.2\textheight,width=0.4\textwidth]
{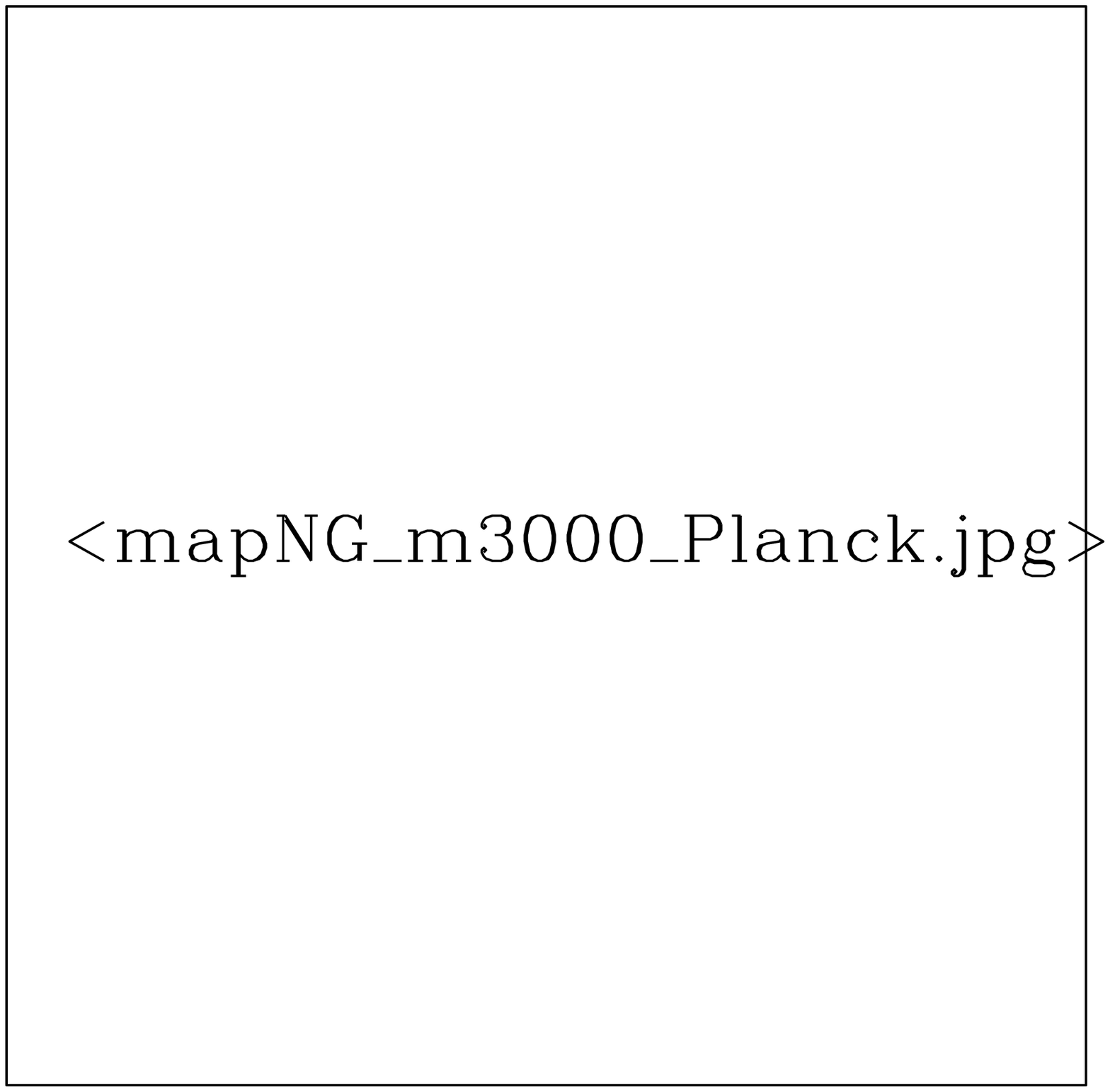}}

\caption{In this panel we show the same Gaussian maps of  
Fig. \ref{fig:maps} on the left-hand side, while on the right-hand side we 
take $f_{NL}=-3000$. The remaining parameters are the same as in Fig. 
\ref{fig:maps}}
\label{fig:maps2}    

\end{figure*}

In Figure 
\ref{fig:pdf} we plot the PDF of temperature fluctuations, obtained 
from an original simulation with $\ell_{\rm max} = 3000$, 
mainly for a comparison with the results contained in (Komatsu {\em et al.} 
2003).  
In Figures \ref{fig:maps},\ref{fig:maps2} we show maps 
obtained from the same simulation, smoothed with different Gaussian beams 
at the resolution of {\em{COBE}} DMR, {\em{WMAP}} and {\em{Planck}}. 

\begin{table}[!b]
\small
\begin{center}
\begin{tabular}{|c|c|c|c|}
\hline
$\mathbf{\ell_{\rm max}}$ & {\bf CPU time} & {\bf Harmonic transform} & 
{\bf nside} \\
\hline
$300$ & $8$ min & $6$ min & $128$ \\
\hline
$500$ & $40$ min & $32$ min & $256$ \\
\hline
$750$ & $1$ h $8$ min & $1$ h  & $256$ \\
\hline
$1500$ & $8$ h $47$ min & $8$ h $20$  min & $512$ \\
\hline
$3000$ & $68$ h $45$ min & $67$ h $15$ min & $1024$ \\
\hline
\end{tabular}
\caption{\rm CPU time needed to simulate a single map}
\label{tab:times} 
\end{center}   
\end{table}

The main purpose of obtaining simulated non-Gaussian CMB maps is that of 
providing a test-bed for the power of different statistical estimators, 
specifically designed to search for non-Gaussianity, in detecting the kind 
and the level of non-Gaussianity which is present in the maps. To this aim, 
all the possible extra effects, such as foreground contamination and 
instrumental noise of a specific experiment, which appear in the statistical 
analysis of real datasets should be properly taken into account.  
These aspects will be the analyzed in a forthcoming paper. 

As already mentioned, there are various non-primordial 
cosmological effects which also imply a deviation from the Gaussian 
behaviour. These include systematic effects as well as some secondary 
sources of anisotropy. Among the future developments of the present 
algorithm is the possibility to include suitable kernels to 
account for some of these effects in the simulated CMB maps. 
This issue will be discussed elsewhere.

\appendix

\section{Derivation of useful formulae}

\subsection{Relation between $\Phi_{\ell m}(k)$ and $\Phi_{\ell m}(r)$}

We start from the Fourier transform
\begin{equation} 
\Phi(\mathbf{r}) = \int \!\frac{d^3k}{(2\pi)^3} 
\,\Phi(\mathbf{k})\,e^{-i\mathbf{k \cdot r}}
\; 
\end{equation} 
and we Rayleigh expand $\exp({-i\mathbf{k \cdot r})}$ in 
spherical harmonics, 
\begin{equation}
\label{eqn:planewaves} 
\exp [-i\mathbf{k}\mathbf{\cdot r}] =
4\pi\sum_{\ell}\sum_{m}(i)^{-\ell}Y_{\ell m}(\mathbf{\hat{k}})Y_{\ell
m}^*(\mathbf{\hat{r}})j_\ell(k r)  \;;
\end{equation} 
we find: 
\begin{eqnarray} 
\Phi(\mathbf{r})& = & \frac{4\pi}{(2\pi)^3}
\sum_{\ell}\sum_{m} (-i)^{\ell}
Y_{\ell m}^*(\mathbf{\hat{r}})\int \! dk\,k^2\,j_{\ell}(kr) \times 
\nonumber \\ & & 
\;\;\;\;\;\;\;\;\;\;\; \times \int \!d\Omega_{\mathbf{\hat{k}}} 
\,Y_{\ell m}(\mathbf{\hat{k}})
\,\Phi(\mathbf{k}) \; . 
\end{eqnarray} 
Recalling that: 
\begin{equation} 
\Phi_{\ell m}(r) = \int \! d\Omega_{\hat{r}} \,
\Phi(\mathbf{r}) \, Y_{\ell m}^{*}(\hat{r}) \;   
\end{equation} 
we can replace the previous expression for $\Phi(\mathbf{r})$ in the last
formula [definition of $\Phi_{\ell m}(r)$] and expand also
$\Phi(\mathbf{k})$ in spherical harmonics. 
We obtain 
\begin{eqnarray}
\label{eqn:s1} 
\Phi_{\ell m}(r) & = & \frac{1}{2\pi^2}
\sum_{\ell_1\ell_2}\sum_{m_1m_2} (-i)^{\ell_1} \!\!\int \!
d\Omega_{\mathbf{\hat{r}}}\, Y_{\ell m}^*(\mathbf{\hat{r}})\,
Y_{\ell_1m_1}^*(\mathbf{\hat{r}}) \times 
\nonumber \\ & &
\!\!\!\!\!\!\!\!\!\!\!\!\!\!\!\times \int \! dk\,k^2
\,j_{\ell_1}(kr)\,\Phi_{\ell_2 m_2}(k)  \int \!
d\Omega_{\mathbf{\hat{k}}}
\,Y_{\ell_1m_1}(\mathbf{\hat{k}})\,Y_{\ell_2m_2}(\mathbf{\hat{k}})  \; . 
\end{eqnarray} 
It is now easy to see, by the orthonormality of spherical
harmonics, that this reduces to Eq. (\ref{eqn:ak2ar}).

\subsection{Correlation function of $\Phi^{\rm L}_{\ell m}(k)$}

Using the definition of $\Phi^{\rm L}_{\ell m}(k)$, we can write
\begin{eqnarray}
\left\langle \Phi^{\rm L}_{\ell_1 m_1}(k_1) 
\Phi^{{\rm L}\star}_{\ell_2 m_2}(k_2) 
\right\rangle & = & \int \!
d\Omega_{\mathbf{\hat{k}_1}}d\Omega_{\mathbf{\hat{k}_2}}\,\left\langle
\Phi^{\rm L}(\mathbf{k_1})\Phi^{{\rm L}\star}
(\mathbf{k_2}) \right\rangle \, \times 
\nonumber \\ & & \;\;\;\;\; \times \; 
Y_{\ell_1 m_1}^\star(\mathbf{\hat{k}_1}) \; 
Y_{\ell_2 m_2}(\mathbf{\hat{k_2}})  \; .  
\end{eqnarray} 
From this we obtain:  
\begin{eqnarray} 
\left\langle \Phi^{\rm L}_{\ell_1 m_1}(k_1) \Phi^{{\rm L}\star}_{\ell_2
m_2}(k_2) \right\rangle & = & (2\pi)^3P_{\Phi}(k_1)\,\int
\!d\Omega_{\mathbf{\hat{k}_1}}d\Omega_{\mathbf{\hat{k}_2}}
\,\delta^D(\mathbf{k_1-k_2}) \times \nonumber \\ & &
\;\;\;\;\;\;\;\;\;\;\; \times \; Y_{\ell_1
m_1}^*(\mathbf{\hat{k_1}})  Y_{\ell_2
m_2}(\mathbf{\hat{k}_2}) \; . 
\end{eqnarray} 
Where we have used: 
\begin{equation} 
\left\langle \Phi^{\rm L}(\mathbf{k_1})
\Phi^{{\rm L}\star}_L(\mathbf{k_2})
\right\rangle = (2\pi)^3 P_{\Phi}(k_1)\delta^D(\mathbf{k_1-k_2}) \; .
\end{equation} 
Recalling now the integral representation of Dirac's delta function, 
\begin{equation}
\delta^D(\mathbf{k_1-k_2}) = \frac{1}{(2\pi)^3} \int \! d^3\mathbf{r} \,\exp
[i(\mathbf{k_1-k_2})\,\mathbf{\cdot r}] \; 
\end{equation} 
we can Rayleigh expand to obtain:\\
$\left\langle \Phi^{\rm L}_{\ell_1m_1}(k_1) 
\Phi^{{\rm L}\star}_{\ell_2m_2}(k_2) \right\rangle = $ 
\begin{eqnarray} && =
16\pi^2P_{\Phi}(k_1)\sum_{\ell_1\ell_2\ell_3\ell_4}
\sum_{m_1m_2m_3m_4} i^{\ell_3-\ell_4}\int drr^2j_{\ell_3}(k_1r)j_{\ell_4}
(k_2r) \times 
\nonumber \\ &&
\;\;\;\;\;\;\;\;  \times \int \!d\Omega_{\mathbf{\hat{r}}}
\,Y_{\ell_3m_3}(\mathbf{\hat{r}})\,Y_{\ell_4m_4}^*(\mathbf{\hat{r}})
\,\int \!d\Omega_{\mathbf{\hat{k}_1}}
\,Y_{\ell_1m_1}^*(\mathbf{\hat{k}_1})\,Y_{\ell_3m_3}^*(\mathbf{\hat{k}_1})
\times 
\nonumber \\ && 
\;\;\;\;\;\;\;\; \times \int
\!d\Omega_{\mathbf{\hat{k}_2}}
\,Y_{\ell_2m_2}(\mathbf{\hat{k}_2})\,Y_{\ell_4m_4}(\mathbf{\hat{k}_2}) \;. 
\end{eqnarray} 
By the orthonormality of spherical harmonics this reduces to 
\begin{eqnarray} 
\left\langle \Phi^{\rm L}_{\ell_1m_1}(k_1) 
\Phi^{{\rm L}\star}_{\ell_2m_2}(k_2)
\right\rangle & = & 16\pi^2P_{\Phi}(k_1)\delta_{\ell_1\ell_2}\delta_{m_1m_2} 
\times \nonumber \\ & &
\!\!\!\!\!\!\!\!\!\!\times \int \!dr\,r^2\,j_{\ell_1}(k_1r)\,j_{\ell_2}(k_2r) 
\; , 
\end{eqnarray}
from which, recalling that 
\begin{equation}
\label{eqn:completezzajl}
\int_0^{\infty} \! dr\,r^2 \,j_\ell(k_1r)\,j_\ell(k_2r) =
\frac{\pi}{2}\frac{1}{k_1^2}\delta^D(k_1-k_2) \; , 
\end{equation} 
we finally obtain 
\begin{equation} 
\left \langle \Phi^L_{\ell_1
m_1}(k_1)\Phi^{L*}_{\ell_2 m_2}(k_2) \right \rangle = 8\pi^3
\frac{P_\Phi(k_1)}{k_1^2} \delta^D(k_1-k_2)\delta_{\ell_1}^{\ell_2}
\delta_{m_1}^{m_2} \; . 
\end{equation}

\subsection{Correlation function of $\Phi^L_{\ell m}(r)$}

According to the Wiener-Khintchine theorem,  
\begin{equation}
\label{eqn:WienerKhintchine} 
\xi_\Phi(r) = \int \!
\frac{d^3k}{(2\pi)^3}\,P_\Phi(k)\,e^{i \mathbf{k} \cdot (\mathbf{r_1 -
r_2})} \; , 
\end{equation} 
where $r = |\mathbf{r_1-r_2}|$, $\xi_\Phi(r)$ and $P_\Phi(k)$ are the
correlation function and the power spectrum of primordial potential
fluctuations, respectively. Expanding once again the exponential in 
spherical harmonics, we obtain 
\begin{equation}
\label{eqn:A} 
\xi_\Phi(r) =
\frac{2}{\pi}\sum_{\ell=0}^{\infty}\sum_{m=-\ell}^{\ell}Y_{\ell
m}(\hat{r}_1)\,Y_{\ell m}^{*}(\hat{r}_2)\,\int \!
dk\,k^2\,P_\Phi(k)\,j_{\ell}(kr_1)j_{\ell}(kr_2) \; .  
\end{equation} 
By the definition of $\Phi_{\ell m}(r)$, we can write  
\begin{equation}
\left\langle \Phi^{L*}_{\ell_1 m_1}(r_1)\Phi^L_{\ell_2 m_2}(r_2)
\right\rangle = \int\!d\Omega_{\hat{r}_{1}}d\Omega_{\hat{r}_{2}}
Y^{*}_{\ell_1 m_1}(\hat{r}_1)\,Y_{\ell_2 m_2}(\hat{r}_2)\,\xi_\Phi(r) \; . 
\end{equation} 
The final result is obtained by substituting 
$\xi_\Phi(r)$ by its expression (\ref{eqn:A}) and using 
the orthonormality of spherical harmonics,   
\begin{equation} 
\left\langle
\Phi^{L*}_{\ell_1 m_1}(r_1)\Phi^L_{\ell_2 m_2}(r_2) \right\rangle =
\frac{2}{\pi}\,\delta_{\ell_1}^{\ell_2}\,\delta_{m_1}^{m_2} \int \!
dk\,k^2\,P_{\Phi}(k) j_{\ell_1}(kr_1)j_{\ell_2}(kr_2) \; . 
\end{equation}

\subsection{Derivation of $W_\ell(r,r_1)$}

Consider a Gaussian random field $\Phi^{\rm L}(\mathbf{r})$ with correlation
function $\xi_\Phi(r)$. If $P_\Phi(k)$ is the corresponding power
spectrum, then $\Phi(\mathbf{r})$ can be obtained as:  
\begin{equation}
\Phi^{\rm L}(\mathbf{r}) = \int \!d^3r_1
n(\mathbf{r_1})\int\!\frac{d^3k}{(2\pi)^3}\sqrt{P_\Phi(k)}
e^{i\mathbf{k}\cdot(\mathbf{r-r_1})}\; , 
\end{equation} 
where we have introduced the white-noise field
$n(\mathbf{r_1})$:  
\begin{equation} 
\left\langle
n(\mathbf{r_1})n(\mathbf{r_2}) \right\rangle =
\delta^D(\mathbf{r_1 - r_2}) \; .  
\end{equation} 
To find a similar expression for $\Phi^{\rm L}_{\ell m}(r)$ we define the 
quantities   
\begin{equation} 
W(\mathbf{r-r_1}) \equiv \int \! 
\frac{d^3k}{(2\pi)^3} \sqrt{P_\Phi(k)}
e^{i\mathbf{k}\cdot(\mathbf{r-r_1})} \; ;  
\end{equation} 
now, as usual, we expand both $\Phi(\mathbf{r_1})$ and $W(\mathbf{r-r_1})$ 
in spherical harmonics and account for the orthonormality relations, 
to finally obtain
\begin{eqnarray} 
\Phi^{\rm L}_{\ell m}(r) & = & \int \! dr_1 \,
r_1^2 \, n_{\ell m}(r_1) W_\ell(r,r_1) \\
W_\ell(r,r_1) & = & \frac{2}{\pi}
\int \! dk \, k^2 \, \sqrt{P_\Phi(k)} \, j_\ell(kr) j_\ell(kr_1) 
\;   
\end{eqnarray}
and 
\begin{equation}
\left
\langle n_{\ell_1 m_1}(r_1)n^\star_{\ell_2 m_2}(r_2) \right \rangle = 
\frac{\delta^D(r_1-r_2)}{r^2}\delta_{\ell_1}^{\ell_2} \delta_{m_1}^{m_2}
\;.  
\end{equation}

\acknowledgements{We acknowledge K. G\'orski for kindly providing 
us with the HEALPix package. We would also like to thank F. Hansen, 
and D. Marinucci for useful discussions.}

\end{document}